\documentclass[aps,floats,twocolumn,showpacs]{revtex4-1}
\usepackage[cp1251]{inputenc}
\usepackage[english]{babel}
\usepackage{amssymb}
\usepackage{amsmath}
\usepackage[pdftex]{graphicx}
\usepackage{cmap}
\usepackage[pdftex]{hyperref}
 \def\bc{\begin{center}}          \def\ec{\end{center}}
 \allowdisplaybreaks

\begin{document}
 \title{ Second harmonic electromagnetic emission in a beam-driven plasma antenna
}
 \author{V.V.Annenkov, E.A.Berendeev, E.P.Volchok, I.V.Timofeev}
 \affiliation{Budker Institute of Nuclear Physics SB RAS, 630090, Novosibirsk, Russia \\
 Novosibirsk State University, 630090, Novosibirsk, Russia \\
Institute of Computational Mathematics and Mathematical Geophysics, 630090, Novosibirsk, Russia}
% \date{\today}
 \begin{abstract}

Generation of electromagnetic radiation near the second harmonic of the plasma frequency during the injection of an electron beam into a rippled-density plasma channel is investigated using both analytical theory and particle-in-cell simulations. The generating scheme is based on nonlinear interaction of the most unstable beam-driven potential plasma wave with its satellite arising due to scattering on the longitudinal modulation of plasma density. Resulting superluminal oscillations of electric current in a finite-size plasma channel radiate electromagnetic waves via the same mechanism which has been recently studied for the fundamental harmonic emission and reffered as a beam-driven plasma antenna. It is shown that theoretical predictions for the optimal plasma width and modulation period are confirmed by simulation results and the power conversion efficiency of the second harmonic emission reaches several percent. Such an efficient mechanism opens the path to explanation of laboratory experiments with a thin electron beam at the GOL-3 mirror trap as well as to developing the scheme of terahertz generation at the gigawatt power level.

 \end{abstract}
 
 \maketitle

\section{Introduction}

Generation of electromagnetic (EM) waves near harmonics of the plasma frequency $\omega_p$ during collective beam-plasma interaction is a well known phenomena that is frequently observed in both space and laboratory plasmas. Presently, processes of linear and nonlinear conversion of beam-driven electrostatic plasma waves  into EM radiation are actively studied in relation to solar radio bursts \cite{Ratcliffe2014,Thurgood2015,Thurgood2016,Che2017} and the problem of high-power terahertz generation \cite{Arzhannikov2012,Arzhannikov2014,Arzhannikov2016,Malik2017,Yang2017}. If the beam-to-radiation power conversion in the latter problem is increased up to a few percent, radiation sources based on multi-GW electron beams will be able to achieve the record gigawatt level of terahertz power.

The principal possibility to reach such a high radiation efficiency (1-2\% near the second harmonic) in a beam-plasma system has been recently demonstrated in laboratory experiments with a 100 keV, 10 MW electron beam at the GOL-3 mirror trap \cite{Burdakov2013,Thumm2014,Ivanov2015}. The main feature of these experiments is a small transverse size of the electron beam which has been chosen comparable with the radiation wavelength. Subsequent theoretical analysis of this regime \cite{Timofeev2015} has shown that such a thin system can radiate EM waves in vacuum as a dipole antenna, if the dominant beam-driven plasma wave with the frequency $\omega_b\approx \omega_p$ generates superluminal satellites via scattering on the periodic density perturbation of plasma ions with the wavenumber $q$:
\begin{equation}
	(\omega_b, k_{\|})+(0,q)\rightarrow (\omega_b, k_{\|}-q).
\end{equation}
We have referred this mechanism as a beam-driven plasma antenna and proposed a theoretical model \cite{Annenkov2016a} allowing to calculate the radiation power at the plasma frequency in the most simple case, when the period of plasma density modulation $2\pi/q$ coincides with the wavelength of the most unstable beam-driven mode $2\pi/k_{\|}$. By changing the ratio $q/k_{\|}$, one can change the angle of the fundamental harmonic radiation.  Generalization of this theory to arbitrary angles \cite{Timofeev2016} has allowed to find the regime in which efficiency of $\omega_p$-antenna radiation is substantially enhanced even in a thick magnetized plasma due to the effect of plasma transparency to radiated EM waves lying slightly below the cut-off frequency $\omega_p$. Further, these theoretical predictions have been successfully confirmed by particle-in-cell (PIC) simulations \cite{Timofeev2017b}. 

Numerical experiments \cite{Annenkov2016b} on the steady-state injection of an electron beam into a previously created rippled-density plasma channel have shown that the maximal efficiency of $\omega_p$-radiation reaches the level of 10\%. It has been also found that the longitudinal modulation of plasma density required for switching on the antenna radiation can arise even in an initially uniform plasma due to the modulational instability of the dominant beam-driven wave. In the regime of self-consistently growing ion modulations, radiation efficiency has been found to be reduced down to 1\%. Thus, the mechanism of a plasma antenna seems to be the most perspective candidate to explain the high efficiency of EM emission observed in the mentioned beam-plasma experiments at the GOL-3 facility. Theoretical interpretation of these experiments will open the path to the use of more energetic (up to 1-2 MeV) thin electron beams for the efficient generation of more powerful radiation in the THz frequency range.   

Experimental spectra of the observed EM radiation as well as first PIC simulations for typical GOL-3 parameters \cite{Annenkov2016c} have shown that the fundamental and second harmonic emissions in these experiments have comparable integral powers. EM emission at the doubled plasma frequency in a thin beam-plasma system  can be generated by nonlinear interaction of the most unstable beam-driven mode with satellite oscillations arising from scattering of the primary wave on the plasma density modulation:
\begin{equation}
	(\omega_b, k_{\|})+(\omega_b, k_{\|}-q) \rightarrow (2\omega_b, 2k_{\|}-q).
\end{equation}
This nonlinear process has been already observed in PIC simulations \cite{Annenkov2016b}, but it remains unclear how the efficiency of this radiation process depends on different beam and plasma parameters. 

Thus, the aim of this paper is to propose an analytical theory capable of predicting the power of the second harmonic EM emission produced by a beam-driven plasma antenna and to verify the adequacy of such a theory by PIC simulations. The proposed theory will allow us to find the optimal modulation number and plasma width, which  is important not only for interpretation of experiments of interest, but also for constructing new generating schemes for  narrow-band terahertz radiation with the record GW power based on the proposed mechanism.

\section{Theory}

Consider the injection of an electron beam with the velocity $v_b$ and relative density $n_b$ into a cold plasma channel immersed in the uniform longitudinal magnetic field. We consider the two-dimensional problem in the Cartesian plane geometry ($x,z$) in order to compare these results with 2D3V PIC simulations, although it can be easily reformulated for the axially symmetric case. The density of this plasma  is expected to be premodulated with the wavenumber $q$: 
\begin{equation}\label{dn}
	n_e=1+\left(\frac{\delta n}{2} e^{ iqz}+c.c.\right).
\end{equation}
We assume that the two-stream instability growing in such a beam-plasma system is dominated by a longitudinally propagating potential Langmuir wave with the electric field 
\begin{equation}
	E_z=\frac{E_0}{2} e^{i k_{\|}z-i \omega_b t}+c.c.
\end{equation}
This wave gets in the Cherenkov resonance with the beam, that is why, at least at the linear stage of instability, the most unstable mode should have the wavenumber  $k_{\|}=\omega_b/v_b$ and frequency $\omega_b=1-n_b^{1/3}/(2^{4/3}\gamma_b)$, where $\gamma_b=(1-v_b^2)^{-1/2}$ is the relativistic factor of the beam. Here, densities are measured in units of the average plasma density $n_0$, velocities --- in the vacuum speed of light $c$, frequencies --- in the plasma frequency $\omega_p=\sqrt{4 \pi e^2 n_0/m_e}$ ($e$ and $m_e$ are the charge and mass of an electron), wavenumbers --- in $\omega_p/c$ and electromagnetic fields --- in units of $m_e c \omega_p/e$. The guiding magnetic field is determined by the dimensionless electron cyclotron frequency $\Omega$. Oscillations of electric field in the dominant beam-driven wave are also accompanied by velocity and density oscillations of plasma electrons:
\begin{equation}
	v_z=v_0 e^{i k_{\|}z-i \omega_b t}+c.c., \qquad \delta n_e=\delta n_0 e^{i k_{\|}z-i \omega_b t}+c.c.
\end{equation}
The presence of the periodic ion density perturbation (\ref{dn}) results in excitation of a long-wavelength satellite in which various quantities $f=\left\{E_z,v_z,n_e\right\}$ oscillate with the shifted longitudinal wavenumber:
\begin{equation}
	\delta f_1(t,z)=\delta f_1 e^{i (k_{\|}-q)z-i \omega_b t}+c.c.
\end{equation}
Contrary to the primary wave with $\delta f_0 \propto E_0$, amplitudes of this satellite are additionally reduced by a small factor of modulation depth, $\delta f_{1}\propto \delta n E_0$. If satellite oscillations reach superluminal phase velocities ($\omega_b/|k_{\|}-q|>1$), they can be linearly converted to electromagnetic waves  near the plasma frequency via the antenna mechanism \cite{Timofeev2015,Annenkov2016a}.  Nonlinear interaction of the primary wave with the satellite generates perturbations near  the doubled plasma frequency
\begin{equation}
	\delta f^{(2)}_1(t,z)=\delta f^{(2)}_{1} e^{i (2 k_{\|}-q)z-i 2 \omega_b t}+c.c., \quad \delta f^{(2)}_{1}\propto \delta n E_0^2.
\end{equation}
Similar oscillations can be also produced by the linear conversion of the forced oscillations at the second harmonic
\begin{equation}
	\delta f^{(2)}_0(t,z)=\delta f^{(2)}_0 e^{i 2 k_{\|}z-i 2 \omega_b t}+c.c.
\end{equation}
with amplitudes $\delta f^{(2)}_{0}\propto E_0^2$ on the density perturbation $\delta n$.
If the phase velocity of arising oscillations exceeds the speed of light,
\begin{equation}
	\frac{2 \omega_b}{|2 k_{\|}-q|}>1,
\end{equation}
the nonlinear current $\delta j^{2}_{1}$ is able to radiate electromagnetic waves in vacuum. It is seen that there is a gap in the $q$-space ($k_{\|}+\omega_b<q<2 (k_{\|}+\omega_b)$) in which the antenna mechanism can generate EM waves only at the second harmonic of the plasma frequency. This gap is best suited to demonstrate the maximal efficiency of the second harmonic antenna emission.

The amplitude of this radiating current ($j_z=-n_e v_z$) can be found as
\begin{equation}\label{cur}
	j^{(2)}_{1}=- \left[v^{(2)}_{1}+\frac{\delta n^{\ast}}{2} v^{(2)}_0 + \delta n_0 v_{1} +\delta n_{1} v_{0}\right].
\end{equation}
Amplitudes of electron density and velocity oscillations are calculated from the continuity equation and equation of motion:
\begin{equation}
	\frac{\partial n_e}{\partial t}+\frac{\partial}{\partial z}\left(n_e v_z\right)=0, \quad \frac{\partial v_z}{\partial t}+v_z \frac{\partial v_z}{\partial z}= -E_z.
\end{equation}
In the linear approximation ($\delta f_0\propto E_0$), we obtain the amplitudes of the primary wave
\begin{equation}
	v_0=-\frac{i E_0}{2 \omega_b}, \qquad \delta n_0=\frac{k_{\|}}{\omega_b} v_0.
\end{equation}
Satellite oscillations arising via scattering of the primary wave on  the density modulation are characterized by the amplitudes:
\begin{equation}
	v_{1}=-\frac{i E_{1}}{\omega_b}, \quad \delta n_{1}=\frac{k_{\|}-q}{\omega_b}\left(v_{1}+\frac{\delta n^{\ast}}{2}v_0\right),
\end{equation}
where the electric field perturbation is found from the Maxwell equation, $\partial E_z/\partial t=- j_z$, and can be written as
\begin{equation}
	E_{1}=\frac{i}{\omega_b \eta_b}\frac{\delta n^{\ast}}{2} v_0, \quad \eta_b=1-\frac{1}{\omega_b^2}.
\end{equation}
The second order velocity perturbations ($\propto E_0^2$) can be presented in the form:
\begin{align}
&v^{(2)}_0=\frac{k_{\|} v_0^2}{\omega_b} \left(\frac{3}{2 \eta}-1\right), \quad \eta=1-\frac{1}{\omega^2}, \\
&v^{(2)}_{1}=\frac{\delta n^{\ast} v_0^2}{\omega^3 \eta} \left[\frac{3 (2k_{\|}-q)}{\eta_b}-\frac{k_{\|}}{2}\frac{\eta_b}{\eta}\right].
\end{align}
where $\omega=2\omega_b$ is the radiation frequency. Substituting all these results in (\ref{cur}), we derive the final expression for the amplitude of radiating current:
\begin{equation}
	j^{(2)}_{1}=\frac{\delta n^{\ast} E_0^2}{\omega^3} \left[\frac{(2 k_{\|}-q)}{\eta_b}\left(1+\frac{3}{\omega^2 \eta}\right)-\frac{k_{\|}}{2}\frac{\eta_b}{\eta^2}\right].
\end{equation}

\begin{figure*}[htb]
	\centering{\includegraphics[width=450bp]{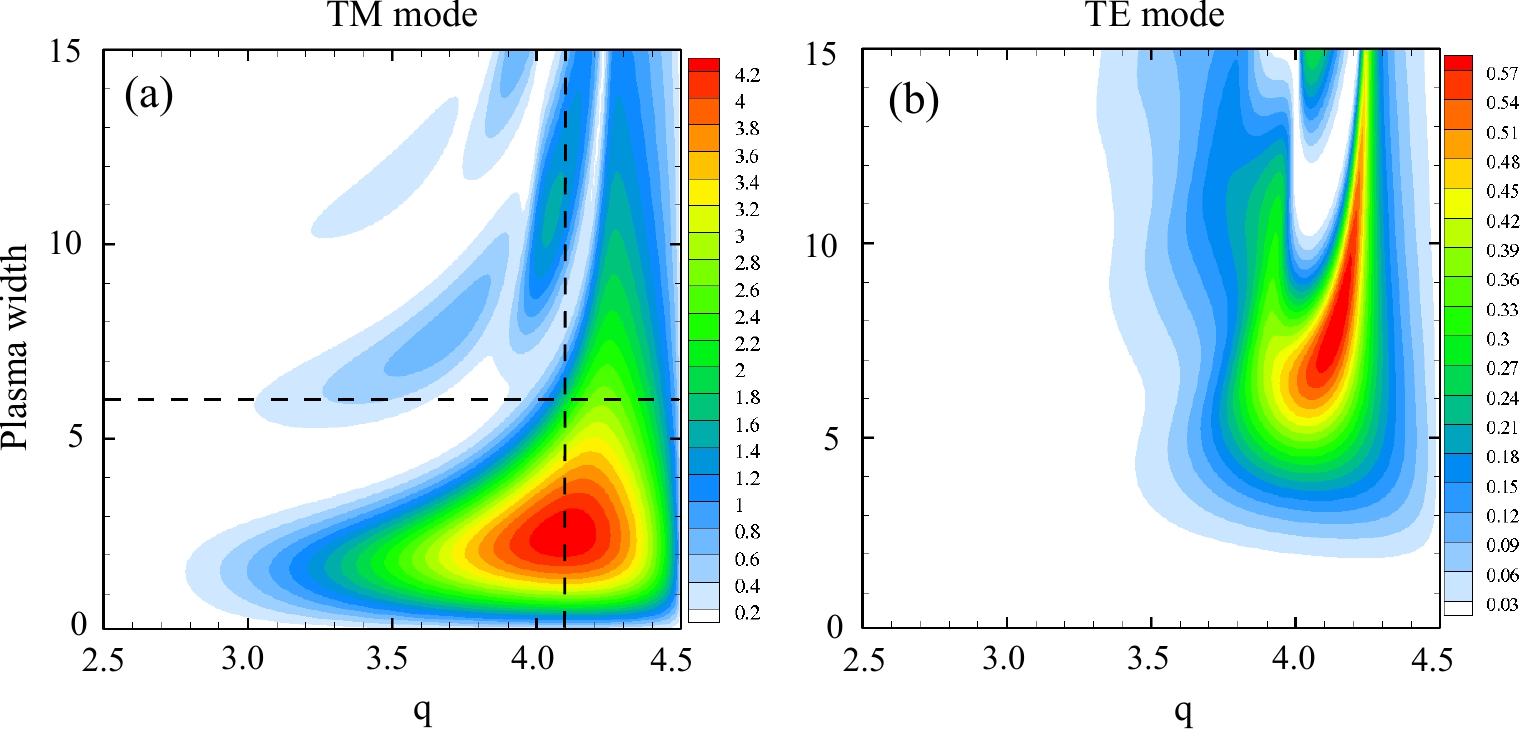}}
	\caption{The radiation efficiency $\mathcal{P}/I$ for TM and TE modes as a function of the modulation wavenumber $q$ (in $\omega_p/c$) and plasma width $2l$ (in $c/\omega_p$) (to obtain real power conversion efficiency $\mathcal{P}$, one should multiply this result by the integral $I=\int|E_0|^4 dz$). The maps are calculated for the case $n_b=0.005$, $v_b=0.9$, $\Omega=0.6$ and $\delta n=0.025$ where the wavenumber of the dominant beam-driven plasma mode reaches the value $k_{\|}=1.25$.}
	\label{fig1}
\end{figure*}
To calculate the radiation power for the given superluminal wave of electric current 
\begin{equation}
	j_z=\mathcal{J} e^{i\mathcal{K}_{\|}z-i \omega t}+c.c.,
\end{equation}
we can use the solution of the Maxwell equations obtained in our recent paper \cite{Timofeev2016} for the fundamental harmonic emission at arbitrary oblique angles. To adapt this solution to our case, we should only change the frequency, longitudinal wavenumber and amplitude of the source current:
\begin{equation}
	\omega=2\omega_b, \quad \mathcal{K}_{\|}=(2k_{\|}-q)/\omega, \quad \mathcal{J}=j^{(2)}_{1}.
\end{equation}
Thus, for the finite plasma width $2l$, efficiency of beam-to-radiation power conversion can be found from the following expression:
\begin{equation}\label{p}
	\mathcal{P}=\frac{2 \mathcal{K}_{\perp} \omega^3 (F_1+F_2)}{(\gamma_b -1) n_b v_b l (\omega^2-1)^2} \int\limits_{0}^{L_z} |\mathcal{J}|^2 dz.
\end{equation}
Here, we separate contributions of different polarizations (TM-mode including $E_x$, $E_z$, $B_y$ fields and TE mode --- $B_x$, $B_z$, $E_y$ fields) to the total radiation power:
\begin{align}
	F_1=&\left|\frac{b_5 \sin(\varkappa_1 l)+G b_6 \sin(\varkappa_2 l)}{Z}\right|^2, \nonumber \\
	F_2=&\left|\frac{b_3 \sin(\varkappa_1 l)+G b_4 \sin(\varkappa_2 l)}{Z}\right|^2,\nonumber 
\end{align}
where $\mathcal{K}_{\bot}=\sqrt{1-\mathcal{K}_{\|}^2}$,
\begin{align}
	Z=b_1& \cos(\varkappa_1 l)+i \mathcal{K}_{\perp} b_5 \sin(\varkappa_1 l)+ \nonumber\\
	&+G\left(b_2 \cos(\varkappa_2 l)+i \mathcal{K}_{\perp} b_6 \sin(\varkappa_2 l)\right),\nonumber \\
	& b_1=a_3-\varkappa^2_1, \quad b_2=i \varkappa_2 a_2, \nonumber \\ 
	& b_3=-i \varkappa_1 a_4, \quad	b_4=a_1-\varkappa_2^2, \nonumber\\
	b_5=& -\varkappa_1 b_1- i \mathcal{K}_{\|} b_3 (\varepsilon -\mathcal{K}_{\|}^2 - \varkappa_1^2)/g,\nonumber \\
	b_6=& -\varkappa_2 b_2- i \mathcal{K}_{\|} b_4 (\varepsilon -\mathcal{K}_{\|}^2 - \varkappa_2^2)/g, \nonumber\\
	G=&-\frac{b_3}{b_4} \left(\frac{\varkappa_1 \cos(\varkappa_1 l)-i \mathcal{K}_{\perp}\sin(\varkappa_1 l)}{\varkappa_2 \cos(\varkappa_2 l)-i \mathcal{K}_{\perp}\sin(\varkappa_2 l)}\right).\nonumber 
\end{align}
TM and TE waves with the given wavenumber $\mathcal{K}_{\|}$ and frequency $\omega$ can propagate inside the plasma as eigenmodes with the transverse wavenumbers
\begin{equation}
	\varkappa^2_{1,2}=\frac{a_1+a_3+a_2 a_4 \mp \sqrt{(a_1+a_3+a_2 a_4)^2-4 a_1 a_3}}{2},
\end{equation}
where
\begin{align}
	&a_1=\eta \left(1-\mathcal{K}_{\|}^2/\varepsilon\right), \quad a_2=\mathcal{K}_{\|}g/\varepsilon, \nonumber \\
	&a_3=\varepsilon-\mathcal{K}_{\|}^2-\frac{g^2}{\varepsilon-\mathcal{K}_{\|}^2}, \quad a_4=\frac{\mathcal{K}_{\|} g}{\varepsilon-\mathcal{K}_{\|}^2}, \nonumber
\end{align}
and
$$\varepsilon=1-\frac{1}{\omega^2-\Omega^2}, \quad g=\frac{\Omega/\omega}{\omega^2-\Omega^2}, \quad \eta=1-\frac{1}{\omega^2} $$ 
are the components of the plasma dielectric tensor.

For the fixed integral $I=\int |E_0|^4 dz$ (assumed independent on $l$ and $q$), our theory predicts that the maximal radiation efficiency $\mathcal{P}$ should be achieved in a thin beam-plasma system ($2l\sim 3 c/\omega_p$) for the emission of EM waves with the TM polarization (Fig. \ref{fig1}). Since the angle of emission is uniquely determined by the wavenumber $q$, 
\begin{equation}\label{ugol}
	\theta=\arctan\left(\frac{\sqrt{\omega^2-(2k_{\|}-q)^2}}{2k_{\|}-q}\right),
\end{equation}
the most efficient radiation should be generated at an acute angle in the backward direction.

The observed position of optimum ($2l\sim 3 c/\omega_p$) can be explained by the fact that only one half-wavelength of a plasma eigenmode should fit the transverse plasma size to provide efficient coupling with the longitudinal radiating current.
In reality, such a small plasma size  influences on the build-up of the beam-plasma instability by reducing the saturation level of plasma oscillations. Thus, the optimal plasma width and modulation period are actually affected by the dependence $I(l,q)$. For this reason, to verify our predictions via PIC simulations, we will further use actual values of the integral  $\int |E_0|^4 dz$ measured directly in the numerical model.

\section{PIC model}
We use the standard 2D3V PIC code with the well known algorithms of Boris and Yee to calculate the self-consistent dynamics of particles and electromagnetic fields, and Esirkepov's scheme to calculate currents. To provide the continuous inflow of particles in a system, we implement open boundary conditions by using special buffers capable of supporting beam and plasma momentum distributions. The description of corresponding algorithms can be found in Ref. \cite{Berendeev2018a}. 
\begin{figure}[htb]
	\centering{\includegraphics[width=240bp]{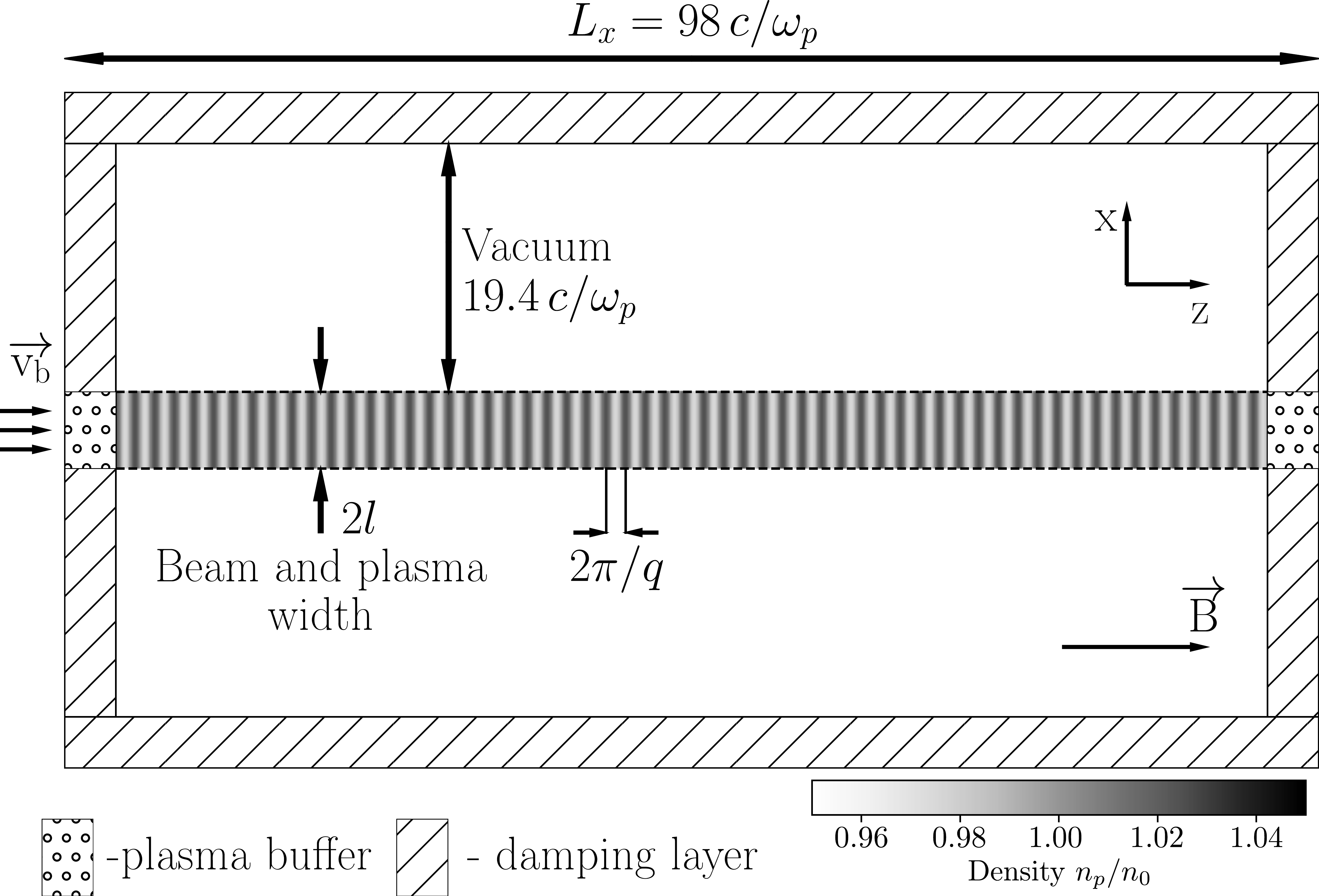}}
	\caption{The layout of a simulation box.}
	\label{fig2}
\end{figure}

The layout of our simulation box is shown in Fig.~\ref{fig2}.
The central part of this box is initially occupied by a plasma in which electrons and ions have nonuniform densities with the periodically perturbed profiles $n_e(z)=n_i(z)=1+\delta n \cos(qz)$. Electrons are characterized by the Maxwellian momentum distribution $f_{e}\propto \exp(-{\bf p}^2/(2 \Delta p_{e}^2))$ with the corresponding temperature $T_{e}= \Delta p_e^2/(2 m_e)=14$ eV and ions are considered as an immobile background. Simulation starts from the state of completely compensated electric charge and zero EM fields. Then we inject beam particles from the left buffer with the shifted Maxwellian momentum distribution corresponding to the relative density $n_b=0.005$, mean velocity $v_b=0.9$ and temperature $T_b=100$ eV. The plasma column is confined by the constant and uniform magnetic field $\mathbf{B}=(0,0,B)$ that is determined by the electron cyclotron frequency $\Omega= eB/(m_e c \omega_p)= 0.6$. 

To measure the power of the generated EM radiation, we use boundary damping layers inside which electromagnetic field values at each time step are multiplied by a coefficient $k<1$ depending on the distance to the boundary \cite{Berendeev2018b}. By using these absorbing layers, we can separate contributions of TM and TE modes to the total radiation power. In the presented simulations, we use the spatial grid $h_z=h_x=0.05$ with the time step $\tau=0.025$ and 144 macroparticles with the parabolic form-factor in a cell for each sort.  

\section{Simulation results}
To verify our theory, let us carry out PIC simulations of the steady-state beam injection into a rippled-density plasma modulated with the depth $\delta n=2.5\%$ for various plasma widths $2l$ and modulation wavenumbers $q$. 

\begin{figure}[htb]
	\centering{\includegraphics[width=240bp]{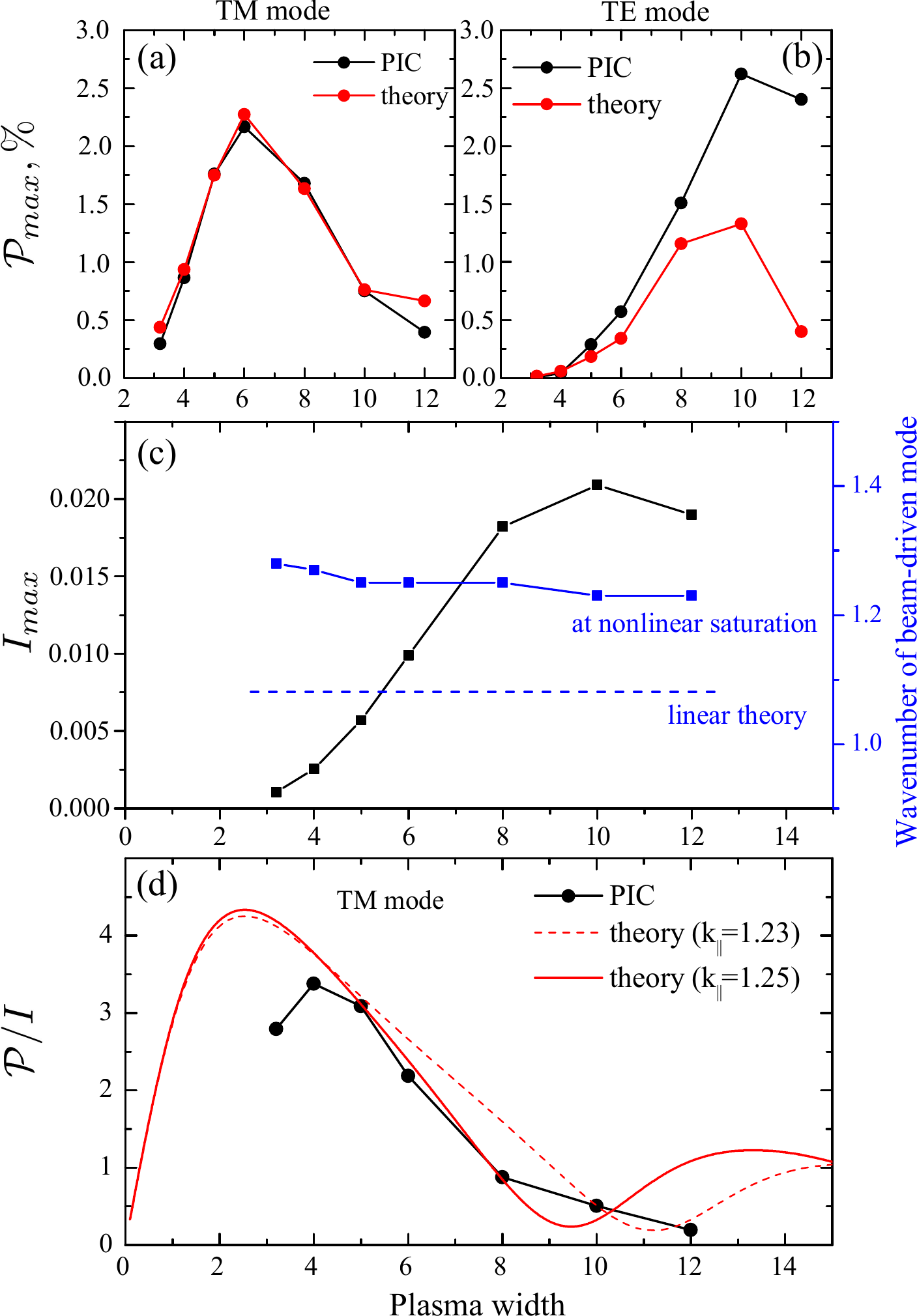}}
	\caption{The maximal efficiency $\mathcal{P}_{max}$ of (a) TM and (b) TE radiations as a function of the plasma width $2l$ for the fixed modulation wavenumber $q=4.1$ in PIC simulations and in theory accounting for the dependence $I(l)$. (c) The dependence of the integral $I_{max}$ and wavenumber $k_{\|}$ on the plasma width observed in PIC simulations. (d) Comparison of theoretical and simulation results in terms of $\mathcal{P}/I$ functions.}
	\label{fig3}
\end{figure}

Let us first study how the relative radiation power $\mathcal{P}$ depends on the plasma channel size for the fixed value $q=4.1$. According to our theory (see the vertical dashed line in Fig. \ref{fig1}a), this modulation wavenumber should be optimal for EM emission. From Fig. \ref{fig3}a presenting the maximal radiation power $\mathcal{P}_{max}$ observed in PIC simulations at each transverse plasma size, one can see that the most efficient radiation of TM-polarized waves  is achieved in a wider plasma ($2l=6$). 
In this regime, more than 2\% of the beam power is converted into the power of TM radiation. It agrees well with the theory if we take into account the strong dependence of the beam-driven plasma wave amplitude $E_0$ and corresponding value of integral $I$ on the plasma width. This dependence is shown in Fig. \ref{fig3}c by the black curve. The beam-plasma instability is seen to be substantially weakened and saturated at lower levels if the plasma size becomes smaller than the wavelength of the dominant unstable mode $\lambda=2 \pi v_b$.  To explain simulation results, we also take into account the nonlinear shift of the longitudinal wavenumber $k_{\|}$ (relative to the position of the Cherenkov resonance with an unperturbed beam $k_{\|}=\omega_b/v_b$) which arises for the most unstable plasma wave at the nonlinear stage of beam trapping. To obtain correct angles of generated radiation, this wavenumber should lie in the close vicinity of $k_{\|}=1.25$.  The appearance of such a shift is also confirmed by the Fourier transform of plasma fields and agrees with our previous simulations \cite{Timofeev2017b}. Thus, in order to exclude variations of radiation efficiency caused by variable amplitudes of plasma waves, it is more suitable to analyze agreement between theory and simulations by comparing the quantities $\mathcal{P}/I$. The theoretical dependence of this parameter on $l$ for the fixed wavenumber $k_{\|}=1.25$ shown in Fig.~\ref{fig3}d is well reproduced by PIC simulations in the wide range of plasma thickness. In the thick plasma channel ($2l=12$), the monotonic reduction of the radiation efficiency $\mathcal{P}/I$ in the PIC model contrasts with the local growth predicted theoretically and can be explained by a small decrease of $k_{\|}$ that strongly affects the radiation intensity at such a plasma width. The efficiency of TE-polarized radiation shown in Fig. \ref{fig3}b reaches the maximum at the greater plasma thickness $2l=10$ in a good agreement with the theoretical prediction, but the theory fails to predict the absolute value of radiation power in this case. The similar excess of the simulated TE radiation power over the theoretical prediction in a beam-driven plasma antenna has been also observed for the fundamental harmonic emission \cite{Timofeev2017b}. It is apparently explained by the greater sensitivity of TE-polarized waves to transverse currents caused by the nonuniform transverse structure of the dominant beam-driven mode. In the theory, contributions of these currents to the total radiation power are not taken into account.

Let us also make sure that the efficiency of the second harmonic emission is really peaked  near the chosen wavenumber $q=4.1$. For this purpose, we perform simulations with various modulation wavenumbers $q$ at the optimal plasma width $2l=6$ (moving along the horizontal dashed line in Fig. \ref{fig1}a). 
\begin{figure}[htb]
	\centering{\includegraphics[width=240bp]{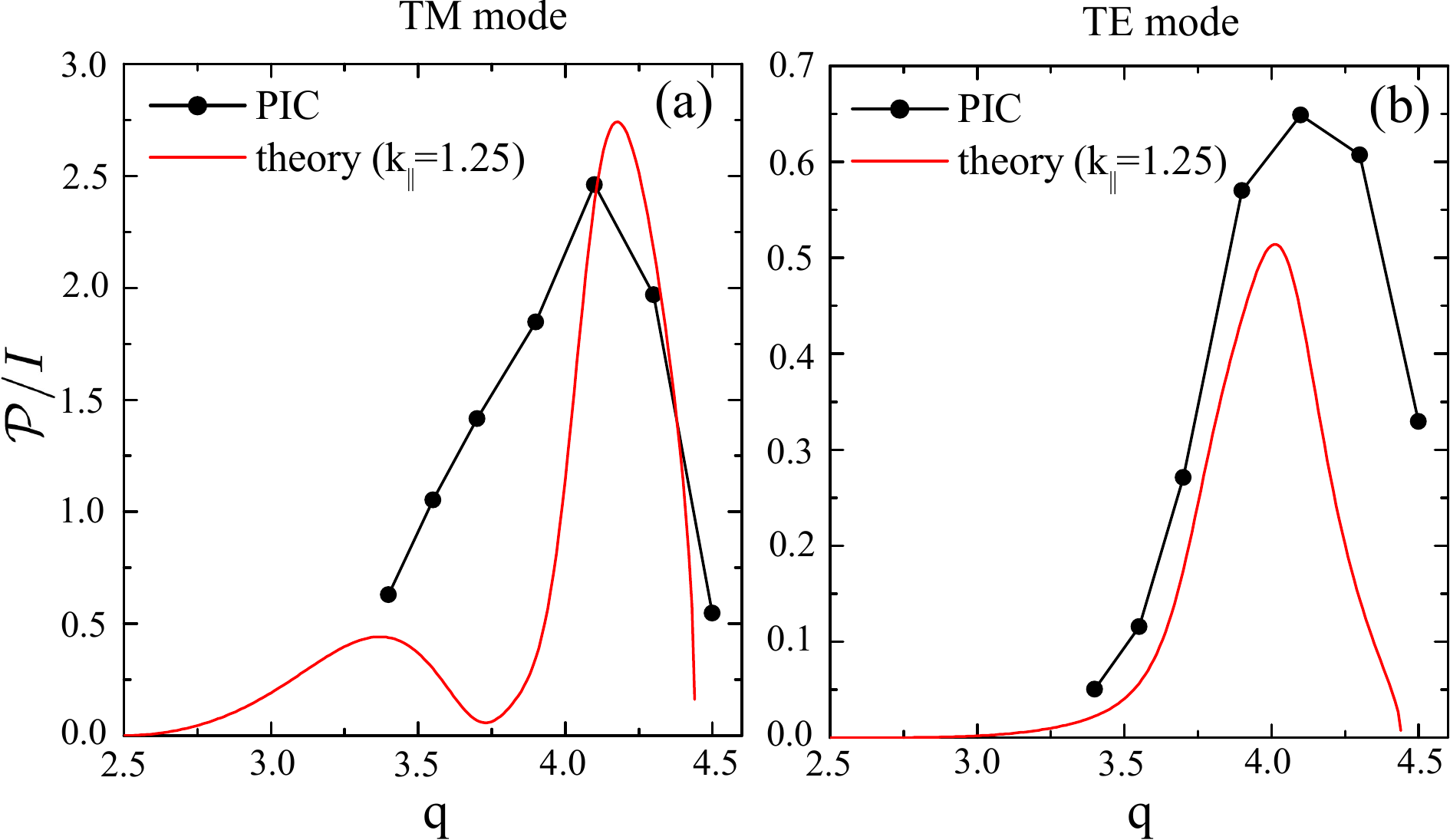}}
	\caption{The dependence of $\mathcal{P}/I$ on the modulation wavenumber $q$ at the plasma width $2l=6$ for TM (a) and TE (b) radiations. Theoretical curves are shown for $k_{\|}=1.25$.}
	\label{fig4}
\end{figure} 
\begin{figure*}[htb]
	\centering{\includegraphics[width=460bp]{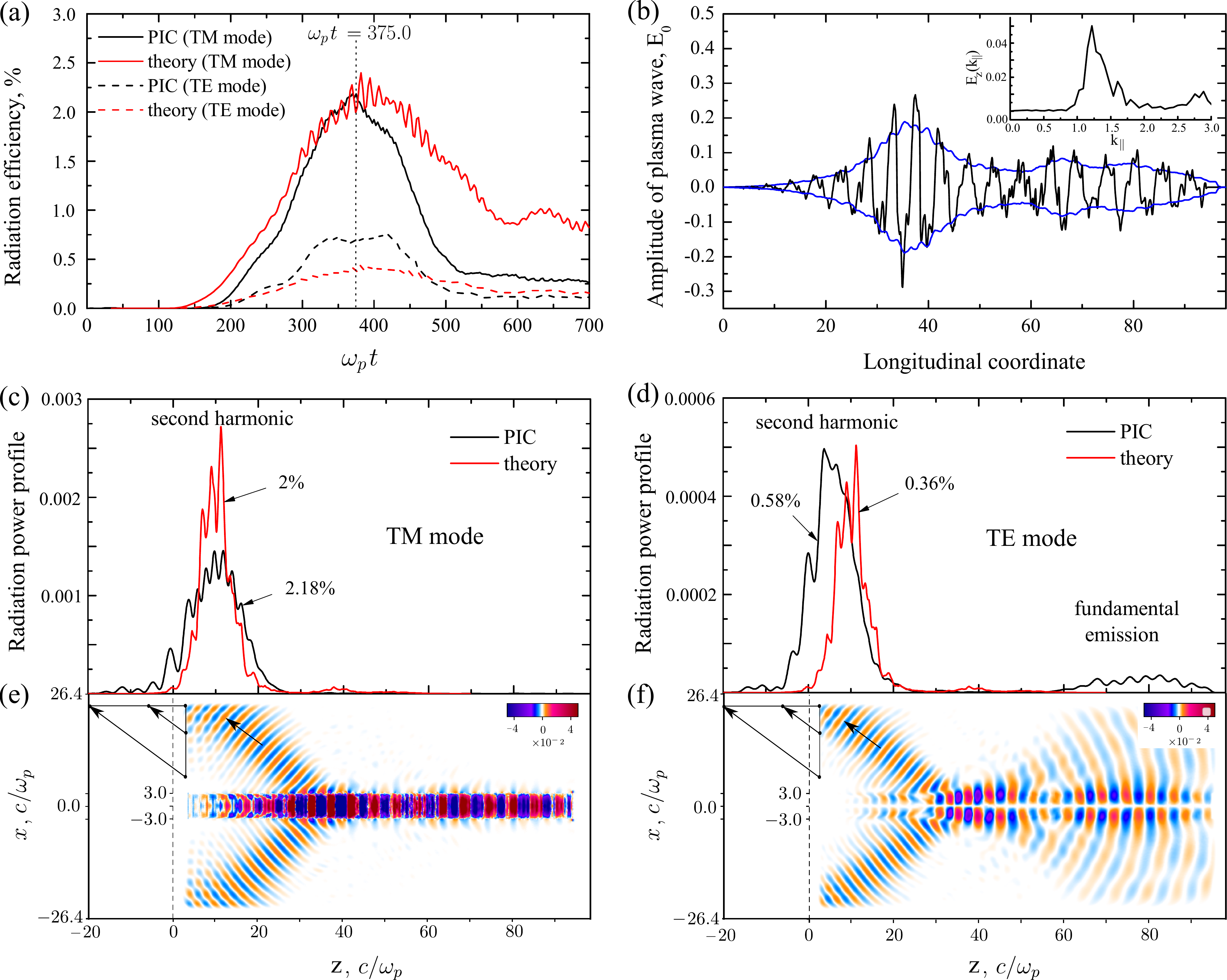}}
	\caption{(a) The history of radiation efficiency $\mathcal{P}$ for TM and TE modes in the case $2l=6$, $q=4.1$. The longitudinal electric field profile $E_z(0,z)$ (black curve) in the central plasma cross section $x=0$ measured in PIC simulations in the moment $t=375$, its Fourier transform (included plot) and the amplitude of plasma wave $E_0$ (\ref{e0}) averaged over the plasma width (blue curve). (c) Simulation and theoretical profiles of radiation power for the TM mode in the same moment of time. (d) Similar profiles for the TE mode. (e) The map of electric field $E_z(x,z)$ (TM mode) in PIC simulations in $t=375$. (f) The map of electric field $E_y(x,z)$ in the same moment of time.}
	\label{fig5}
\end{figure*} 
Fig. \ref{fig4} shows that $q$-dependences of the radiation efficiency for both TM and TE modes are qualitatively reproduced in PIC simulations and reach their maximal values inside the predicted region. The observed broadening of the radiation power profile for TM mode towards small $q$ is apparently associated with the deviation of $k_{\|}$from the value $k_{\|}=1.25$ realized in the optimal $q$-region. Since the integral $I$ remains almost independent on $q$, the real radiation efficiency $\mathcal{P}$ follows the same $q$-profiles.  

To prove the dominance of the antenna mechanism in generating the observed second harmonic  emission, let us consider the most efficient regime ($2l=6$, $q=4.1$) in more details. The temporal evolution of the radiation power $\mathcal{P}$ for both polarizations is shown in Fig. \ref{fig5}a. 
Here, at each time step, we compare the power absorbed in input and side damping layers (in units of the beam power) with the theoretical dependence $\mathcal{P}=T\cdot I$ in which $T$ is calculated from the analytical formula (\ref{p}) and the integral $I=\int |E_{0}|^4 dz$ is directly measured in simulations using real beam-driven fields in a plasma. It is seen that the temporal behavior of the observed radiation power is really determined by the evolution of $I$. To obtain the amplitude $E_0(z)$ of a dominant plasma wave for the given spatial distribution of longitudinal electric field $E_z(x,z)$, we first calculate the local values
\begin{equation}
	E_0(x,z)=\left[\frac{k_0}{\pi}\int\limits_{-\pi/k_0}^{\pi/k_0} dz^{\prime} E_z^2(x,z+z^{\prime})\right]^{1/2} 
\end{equation}
($k_0=\omega_b/v_b$) and then average them over the plasma width
\begin{equation}\label{e0}
	E_0(z)=\left[\frac{1}{2l} \int\limits_{-l}^{l} dx E_0^4(x,z)\right]^{1/4}.
\end{equation}
 The resulting amplitude and the real oscillating field $E_z(0,z)$ in the moment of the most intense radiation $t=325$ is presented in Fig. \ref{fig5}b. The Fourier spectrum of the field $E_z(0,z)$ shows that the longitudinal wavenumber of this nonlinear wave is really shifted to the value $k_{\|}=1.25$. Due to the antenna mechanism (see the formula (\ref{ugol})), the second harmonic radiation in this case should be emitted at the angle $\theta=34^0$. The corresponding direction of radiation wavenumber shown in Fig. \ref{fig5}e and \ref{fig5}f by arrows is found to coincide with the emission direction  observed in PIC simulations. The maps of electric fields $E_z$ and $E_y$ corresponding to TM and TE modes in Fig. \ref{fig5}e and \ref{fig5}f do really demonstrate the dominance of this narrowly directed second harmonic emission, although a relatively small part of energy is also converted to the parasitic TE-polarized fundamental radiation in the forward direction. To make sure that not only integral characteristics, but also spatial profiles of absorbed radiation power are satisfactory predicted by our theory, we compare these profiles with theoretical ones $T|E_0|^4$. Since some part of radiation energy in the PIC model falls in the input wall, we project the absorbed power to the $z$-axis and, after summation of contributions of top and bottom walls, obtain a single profile of radiation power for each polarization (black curves in Fig. \ref{fig5}c and \ref{fig5}d). It is seen that the TM profile (Fig. \ref{fig5}c) is well superposed with theoretical one: it contains the same integral power, but appears to be wider due to the diffraction (not accounted for in the theory). The profile of TE mode (Fig. \ref{fig5}d) is worse described by the theory, since this radiation is surprisingly emitted at slightly different angle (resulting in the observed shift of the central line) and noticeably exceeds the integral theoretical power.

\begin{figure}[h]
	\centering{\includegraphics[width=240bp]{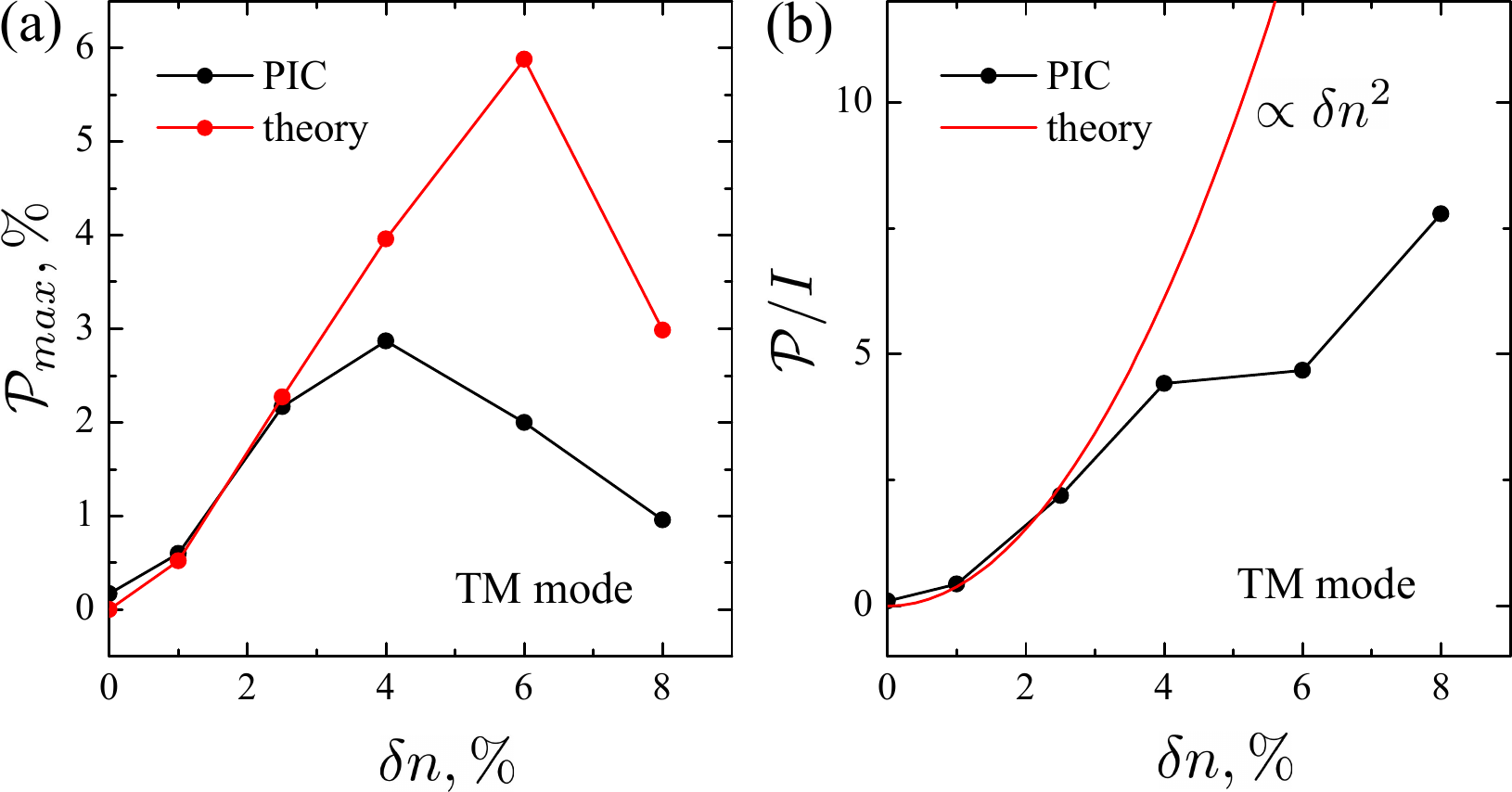}}
	\caption{(a) The maximal efficiency of TM radiation $\mathcal{P}_{max}$ as a function of the modulation depth $\delta n$ (in \%). (b) Comparison of theoretical and simulation results in terms of $\mathcal{P}/I$ functions.}
	\label{fig6}
\end{figure}

Let us also find out whether the radiation efficiency can be increased in a plasma with deeper density modulations. For this purpose, we carry out simulations with various $\delta n$ and fixed $2l=6$ and $q=4.1$. If the modulation depth does not influence on the amplitude of dominant plasma wave ($I$ does not depend on $\delta n$), the radiation efficiency should grow according to the law $\propto \delta n^2$. In reality, the growth of $\delta n$ is accompanied with the substantial decrease of the integral $I$ which appears to be stronger than $\delta n^2$ already in the range $\delta n>4\%$. Figure \ref{fig6}a shows 
that the efficiency of TM-polarized radiation due to the strong dependence $I(\delta n)$  reaches the maximal value 2.6\% for the depth $\delta n=4\%$ and then reduces. By comparing the quantities $\mathcal{P}/I$ (Fig. \ref{fig6}b), we can conclude that the theoretical scaling $\mathcal{P}/I\propto \delta n^2$ actually works only for small modulation depths $\delta n<3\%$. The same range of applicability for the plasma antenna theory has been found in the regime of $\omega_p$-radiation \cite{Timofeev2017b}.

\section{Summary}
We propose a theoretical model of a beam-driven plasma antenna to calculate the power of electromagnetic radiation generated near the doubled plasma frequency during the injection of an electron beam into a magnetized rippled-density plasma channel. Theoretical predictions for the spatial distribution of radiation power as well as for the optimal plasma width and optimal modulation period of plasma density are well reproduced by PIC simulations. It is shown that, due to the antenna mechanism, the efficiency of beam-to-radiation power conversion can reach several percent. It means that, in principal, the mechanism of a plasma antenna is able to explain the highly efficient second harmonic emission observed in experiments with a thin 100 keV electron beam at the GOL-3 facility and can be used to generate narrow-band GW-level THz radiation by multi-GW electron beams capable of focusing into millimeter spots.

\begin{acknowledgments}
This work is supported by the Russian Foundation for Basic Research (grant 18-02-00232). Simulations are carried out using computational resources of Peter the Great Saint-Petersburg Polytechnic University Supercomputing Center (www.spbstu.ru) and Novosibirsk State University. 
\end{acknowledgments}

\end{document}